\documentclass{article}

\usepackage{arxiv}

\usepackage[utf8]{inputenc} 
\usepackage[T1]{fontenc}    
\usepackage{hyperref}       
\usepackage{url}            
\usepackage{booktabs}       
\usepackage{amsfonts}       
\usepackage{nicefrac}       
\usepackage{microtype}      
\usepackage{amsmath}        
\usepackage{lipsum}
\usepackage{graphicx}
\usepackage{algorithm}
\usepackage{algorithmic}
\usepackage{cite}
\graphicspath{ {./images/} }

\title{Z-Space: A Multi-Agent Tool Orchestration Framework for Enterprise-Grade LLM Automation}

\author{
 Qingsong He \\
  Rajax Network Technology (ele.me)\\
   \And
 Jing Nan \\
  Rajax Network Technology (ele.me)\\
    \And
   Jiayu Jiao \\
  Rajax Network Technology (ele.me)\\
  \And
  Liangjie Tang \\
  Rajax Network Technology (ele.me) \\
  \texttt{email} \\
  \And
  Xiaodong Xu \\
  Rajax Network Technology (ele.me) \\
  \And
  Mengmeng Sun \\
  Rajax Network Technology (ele.me) \\
  \texttt{email} \\
  \And
  Qingyao Wang\\
  Rajax Network Technology (ele.me)\\
  \And
  Minghui Yan\\
  Rajax Network Technology (ele.me)\\
  \AND
  \texttt{yinan.nj@alibaba-inc.com}
}

\begin{document}
\maketitle
\begin{abstract}
Large Language Models can break through knowledge and timeliness limitations by invoking external tools within the Model Context Protocol framework to achieve automated execution of complex tasks. However, with the rapid growth of enterprise-scale MCP services, efficiently and accurately matching target functionalities among thousands of heterogeneous tools has become a core challenge restricting system practicality. Existing approaches generally rely on full-prompt injection or static semantic retrieval, facing issues including semantic disconnection between user queries and tool descriptions, context inflation in LLM input, and high inference latency. To address these challenges, this paper proposes Z-Space, a data-generation-oriented multi-agent collaborative tool invocation framework Z-Space. The Z-Space framework establishes a multi-agent collaborative architecture and tool filtering algorithm: (1) A structured semantic understanding of user queries is achieved through an intent parsing model; (2) A tool filtering module (FSWW) based on fused subspace weighted algorithm realizes fine-grained semantic alignment between intents and tools without parameter tuning; (3) An inference execution agent is constructed to support dynamic planning and fault-tolerant execution for multi-step tasks. This framework has been deployed in the Eleme platform's technical division, serving large-scale test data generation scenarios across multiple business units including Taotian, Gaode, and Hema. Production data demonstrates that the system reduces average token consumption in tool inference by 96.26\% while achieving a 92\% tool invocation accuracy rate, significantly enhancing the efficiency and reliability of intelligent test data generation systems.
\end{abstract}


\section{Introduction}
In recent years, Large Language Models have achieved groundbreaking advancements in natural language understanding and generation, demonstrating formidable capabilities in general reasoning and dialogue\cite{thirunavukarasu2023large}. However, their capabilities are fundamentally constrained by the static snapshot of training data and limited context window, making it challenging to directly intervene in real-world state transitions or access private, dynamic data sources\cite{naveed2025comprehensive}. To overcome these limitations, technologies such as Function Calling and Model Context Protocol have emerged, endowing LLMs with the ability to invoke external tools, execute program code, and manipulate physical or digital environments\cite{hou2025model}. This marks a paradigm shift in artificial intelligence systems—from "passive responders" toward "active executors."

Despite these advancements, existing tool integration paradigms face significant challenges. The prevailing mainstream approach adopts an "all-injection" strategy, where detailed descriptions (name, parameters, functional specifications) of all available tools are concatenated into the prompt to support LLM decision-making\cite{danthine2003protocol}. While conceptually simple, this method reveals fundamental flaws in practical applications: As the tool repository scales, prompt length expands linearly. This not only incurs substantial computational costs and latency but also drowns critical semantic cues in massive redundant information, dispersing model attention and significantly degrading decision accuracy—effectively regressing an advanced cognitive agent into an inefficient keyword matching engine.

To alleviate contextual pressure, recent studies have explored Retrieval-Augmented Generation based tool selection methods\cite{arslan2024survey}. These approaches vectorize tool metadata and perform similarity retrieval against user queries at runtime to pre-screen candidate tools\cite{gan2025rag}. While such methods partially address scalability issues, they remain trapped in static, passive retrieval paradigms. Their fatal deficiency lies in the lack of foresight and dynamic adaptation: Systems can only respond to the literal meaning of initial queries, failing to anticipate emergent tool requirements arising from intermediate states during task execution chains\cite{shen2024llm}. For instance, when receiving the instruction "Issue a coupon to the user," an ideal execution path should involve consecutive actions including user information retrieval, coupon distribution, and result verification. However, single-stage retrieval based on the initial query often only identifies the "coupon distribution" module, leaving other required tools unactivated.

Addressing these limitations, this paper proposes the Z-Space framework, whose core contribution lies in constructing a hierarchical, dynamic tool orchestration mechanism with deep semantic comprehension. We abandon coarse-grained full injection and shallow static retrieval, introducing instead a multi-agent architecture spanning intent recognition, tool filtering, reasoning execution, and summary interaction. Notably, the tool filtering phase innovatively incorporates a Fused Subspace with Word Weights (FSWW) algorithm. By simulating attention mechanisms and residual connections to model deep feature interactions, this algorithm achieves fine-grained semantic relationship modeling between intents and tools without gradient-based training. The algorithm not only drastically reduces contextual load but also empowers the system to dynamically focus on critical tools through semantic relevance. Combined with parent-child intent decomposition and asynchronous task orchestration, our framework enables cross-step tool dependency reasoning, realizing a paradigm shift from "passive lookup" to "proactive planning." Experimental validation demonstrates that the system significantly improves complex task success rates and execution efficiency while maintaining low latency, offering novel insights for enterprise-grade intelligent data generation and multi-agent architectures.

\section{Related work}
\subsection{Tool-Integrated Applications of Large Language Models}
To overcome limitations in arithmetic computation, information retrieval, and code execution, researchers have enhanced large language models through integration with external tools\cite{qu2025tool}. Recent years have witnessed significant progress in LLM-based tool orchestration frameworks, primarily focusing on three directions: context optimization, tool retrieval, and task orchestration\cite{zhan2024injecagent}. Early studies predominantly adopted the Full Prompt Injection strategy, directly concatenating tool descriptions (e.g., name, parameters, functional specifications) into prompts\cite{xie2024travelplanner}. For instance, OpenAI's Function Calling mechanism enables tool invocation via structured parameter definitions but struggles with large-scale tool repositories due to context window constraints\cite{bacon1996fast}. To mitigate context inflation, researchers proposed a RAG-based tool retrieval framework leveraging vectorized tool metadata for runtime dynamic retrieval, though its static retrieval paradigm fails to adapt to evolving task requirements during execution\cite{zhao2025tura}.

In dynamic task planning, Microsoft's AutoGen framework achieves task decomposition through multi-agent collaboration but relies on manually defined agent roles and interaction rules, limiting flexibility\cite{wu2024autogen}. Google's SayCan system combines tool invocation with value functions via reinforcement learning for tool selection, yet its high training costs hinder adaptability in rapidly changing enterprise environments\cite{ahn2022can}. Recently, frameworks like LangChain and LlamaIndex attempt to address multi-step task orchestration through predefined execution chains\cite{autry2008supply}. However, their reliance on fixed process templates prevents handling of dynamic dependencies.
\subsection{Retrieval-Augmented Generation}
Research targeting hallucination suppression and factual accuracy enhancement in LLMs has driven breakthroughs in Retrieval-Augmented Generation systems by anchoring generation processes to external knowledge\cite{autry2008supply}. Early studies employed static "retrieval-generation" paradigms, focusing on optimizing coordination between retrieval and generation modules\cite{chen2025cmrag}. As the field evolved, RAG frameworks transitioned from rigid workflows to dynamic adaptive architectures: Self-RAG utilizes "reflection tokens" for demand-driven retrieval decisions, avoiding resource waste from invalid queries\cite{asai2024self}; active RAG adopts iterative retrieval strategies to support dynamic information acquisition during generation; R2AG bridges the semantic gap between retrievers and generators through fine-grained retrieval features\cite{treer}. Nevertheless, commercial systems like Perplexity AI and Google AI Overviews demonstrate that existing solutions still face challenges in content quality maintenance and context comprehension robustness. Their fundamental limitation lies in rigid dependence on predefined workflows, hindering intelligent integration of diverse information sources and multi-tool coordination for complex queries.

Initially, RAG technology combined parametric LLMs with non-parametric memory systems through dense vector indexing, retrieving relevant text passages during inference to enhance knowledge-intensive tasks\cite{selva2021review}. Subsequent research expanded its application to generalized NLP paradigms, including modular advanced variants that dynamically adjust at token/query levels\cite{hupkes2023taxonomy}. This architecture—decoupling memory access from text generation—inspired our MCP-RAG framework, treating MCP discovery as an orthogonal sub-problem to text generation\cite{gan2025rag}. Current research aims to transcend traditional workflows by building intelligent frameworks capable of autonomous retrieval timing judgment and dynamic multi-tool chain coordination, addressing enterprise demands for seamless integration of complex, multi-source information.

Compared to conventional methods, our Z-Space framework achieves breakthroughs across three dimensions: (1) The FSWW algorithm integrates subspace projection with dynamic residual connections to enable fine-grained semantic alignment in an unsupervised manner; (2) An intent tree decomposition mechanism supports dynamic dependency reasoning between parent-child intents; (3) An asynchronous orchestration engine facilitates fault-tolerant recovery and dynamic planning during task execution. Compared to traditional RAG methods, our framework significantly improves tool matching accuracy through semantic enhancement while maintaining low context overhead.

\section{Method}
\begin{figure}
    \centering
    \includegraphics[width=0.5\linewidth]{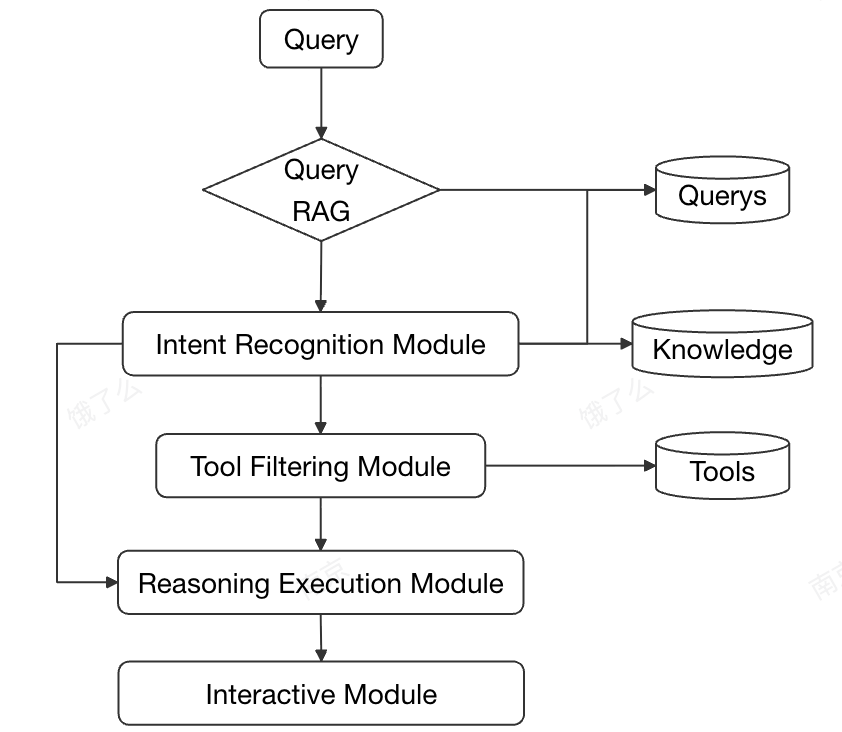}
    \caption{Architecture of Z-Space}
    \label{jiagou}
\end{figure}

This paper proposes a data-generation-oriented multi-agent collaborative tool invocation framework Z-Space, designed to bridge the semantic gap between the general cognitive capabilities of Large Language Models and the functional APIs of enterprise proprietary systems. The framework achieves precise mapping from unstructured user instructions to deterministic tool invocations through a hierarchical and interpretable decision-making mechanism. As illustrated in figure $\ref{jiagou}$, the system architecture follows the "perception-decision-execution" cognitive computing paradigm, comprising four core modules: Intent Recognition, Tool Filtering, Reasoning Execution, and Interactive Module. These modules are interconnected through explicitly defined interfaces for data flow, collectively enabling automated orchestration of complex tasks.

\subsection{Multi-Agent Architecture}
The model architecture adopts an event-driven pipeline framework, transforming user-initiated chat requests into structured responses. As illustrated in Figure \ref{jiagou}, incoming requests first enter the Intent Recognition module, where a prompt engineering-based LLM performs deep semantic parsing of user queries to output a structured intent representation. This representation captures both high-level objectives and low-level operational details of user instructions. Subsequently, the Tool Filtering module receives the structured intent and registered tool metadata pool, generating an ordered candidate tool list through a dual-track matching algorithm that integrates semantic similarity computation with heuristic rules. This process constitutes the critical embodiment of system intelligence. Following this, the Reasoning Execution module—acting as the "Executor"—coordinates actual tool invocations based on candidate tools and potential execution plans. It manages concurrency, error recovery, and result aggregation for tool calls, delivering final results to the Summary Interaction Agent for natural language refinement to produce fluent end-user responses. The entire workflow is implemented on the SpringAI framework, ensuring system scalability and operational efficiency.
\subsection{Intent Recognition Module}
The Intent Recognition Module focuses on intent identification and structured parsing of user queries. The system designs a hierarchical intent parsing model to overcome the inherent ambiguity and polysemy of natural language. At its core lies an intent recognition Agent that performs zero-shot or few-shot learning through predefined system prompts, guiding it to jointly extract multiple semantic features from user queries.

Specifically, the intent recognition Agent identifies the following key semantic elements:

\begin{itemize}
\item Main Intent Category: Represents the macroscopic task category (e.g., "data generation," "query").
\item Operation: Specifies concrete actions (e.g., "create," "query," "update," or "delete").
\item Target Object: Identifies the object instance for the operation (e.g., "product," "coupon").
\item Execution Plan: A sequence of steps (including step descriptions and data dependency relationships) required to fulfill the user's objective.
\end{itemize}

To handle composite tasks, the system supports intent tree construction. A main intent can be recursively decomposed into a series of ordered sub-intents, forming an executable action plan. Additionally, to enhance domain-specific terminology comprehension, the system integrates a semantic normalization component. This component maintains a domain-specific synonym dictionary to normalize extracted semantic slots, effectively mitigating semantic mismatch issues caused by lexical variations and significantly improving intent recognition accuracy.

\subsection{Tool Filtering Module}
The Tool Filtering Module focuses on extracting the most relevant tools from a vast pool of MCP tools. Its core component is the Fused Subspace with Word Weights (FSWW) algorithm, designed to address the critical challenge of effectively injecting and emphasizing local semantic information from specific keywords while preserving the holistic semantic integrity of original statements. The FSWW algorithm achieves efficient and robust semantic enhancement through a multi-stage, adaptive fusion strategy. The following section provides a detailed exposition of the FSWW algorithm design.
\paragraph{Word Weight Computation.}
The first step evaluates the importance of each keyword within the original statement. We employ an attention mechanism, computing the cosine similarity between the normalized statement vector $\bar{A}$ (L2-normalized version of A) and each keyword vector ${w_j}$ as its weight ${\omega_j}$:
\begin{equation}
    \omega_j = \text{Cosine}(\bar{A}, w_j)
\end{equation}

All keyword weights collectively form a weight vector $\Omega = [\omega_1, \omega_2, ... \omega_k]$. This design endows the model with selective attention capability, ensuring that semantic contributions primarily originate from keywords highly correlated with the original statement's thematic core.

\paragraph{Weighted Subspace Projection}
Traditional subspace projection methods treat keywords as equally weighted basis vectors. FSWW improves this approach through weighted subspace projection. This study constructs a weighted basis matrix $S$ by combining keyword vectors with their corresponding weight vector $\Omega$, where the j-th column is defined as $\omega_j \cdot \bar{w}_j $ ($\bar{w}_j$ represents the L2-normalized word vector). Subsequently, the projection of the normalized statement vector $\bar{A}$ onto this weighted subspace is computed as:

\begin{equation}
    P_s(\bar{A}) = S(S^\top S + \epsilon I)^{-1}S^\top\bar{A}
\end{equation}

Here, $\epsilon$ denotes the ridge regression parameter that prevents singularity of the $S$ matrix and improves numerical stability. The essence of this projection operation is to bias $\bar{A}$ towards semantic directions dominated by keywords, with the weight vector $\Omega$ granting greater influence to critical keywords in shaping the subspace structure.

\paragraph{Weighted Semantic Component Extraction}
Beyond subspace projection, the FSWW model extracts and weights keyword semantic information from three distinct perspectives to form complementary fusion signals:

Weighted Word Center Bias (WWCB): Computes the weighted average vector $\bar{E}$ of keywords as their collective semantic "centroid." This component gently biases the fusion result toward the geometric center of the keyword cluster:

\begin{equation}
    \bar{E} = \text{Normalize}\left(\frac{\sum_{j=1}^{k} \omega_j \cdot \bar{w}_j}{\sum_{j=1}^{k} \omega_j}\right)
\end{equation}

Weighted Differential Vector (WDV): A novel innovation in our model. We calculate the differential vector $\Delta_j = \bar{\omega}_j - \bar{A}$ for each keyword $\bar{w}_j$, representing the "semantic increment" introduced by that word. A weighted summation of these differentials yields the desired semantic correction direction $\bar{W}$:

\begin{equation}
    \bar{W} = \text{Normalize}\left(\sum_{j=1}^{k} \omega_j \cdot (\bar{w}_j - \bar{A})\right)
\end{equation}

\paragraph{Multi-Component Linear Fusion}
The components described above are linearly combined with the original statement vector to form the preliminary fusion vector $F$:

\begin{equation}
    F = \text{Normalize}\left[(1-\alpha)\bar{A} + \alpha \cdot P_s(\bar{A}) + \beta \cdot \bar{E} + \gamma \cdot \bar{W}\right]
\end{equation}

Here, $\alpha$, $\beta$, and $\gamma$ are learnable or tunable hyperparameters for subspace projection, word center bias, and differential vector contributions respectively. This equation explicitly models four competing forces: preserving the original statement $(1-\alpha)\bar{A}$, aligning with the keyword subspace $\alpha \cdot P_s (\bar{A)}$, approaching the keyword centroid $\beta \cdot \bar{E}$, and evolving along the correction direction $\gamma \cdot \bar{W}$.

\paragraph{Dynamic Residual Connection and Semantic Protection}
To ensure robustness during fusion, the model incorporates two-layer safety mechanisms: dynamic residual connection and semantic guard mechanism. Inspired by ResNet\cite{he2016deep}, the final output $V$ is a weighted average of the original statement $\bar{A}$ and preliminary fusion result $F$. Unlike static weighting, this study introduces a dynamic gating mechanism where the weight $\Gamma$ is determined by the mean similarity $\mu_{\omega}$:

\begin{equation}
    \Lambda = {\lambda} \times (0.3 + 0.7 \times \mu_\omega)
\end{equation}
\begin{equation}
    V = (1-\Lambda) \times \bar{A} + \Lambda \times \frac{F}{||F||}
\end{equation}
As the final safeguard, the model enforces a requirement that the cosine similarity between the output $V$ and original statement $\bar{A}$ must not fall below a predefined threshold $\Delta$. If $cos(\bar{A},V)<\Delta$, the fusion is deemed excessive. The model then proportionally scales down the fusion coefficients $\alpha$, $\beta$, $\gamma$ and re-executes the fusion process until the similarity constraint is satisfied or the maximum iteration limit is reached.

\subsection{Reasoning Execution Module}
The Reasoning Execution module transforms abstract tool candidate lists into observable action sequences. This module employs an asynchronous task orchestration strategy to maximize system throughput.

The module first executes all auxiliary tool calls concurrently to parallelly acquire prerequisite data. Subsequently, core tools are activated sequentially or in parallel based on parent-child intent dependency relationships derived from intent parsing. Each tool invocation is encapsulated as an independent asynchronous task, scheduled uniformly by a thread pool. Raw data from tool responses are staged and used as inputs for subsequent steps.

Upon completion of all required tool invocations, a top-level LLM agent is triggered to perform the Result Synthesis task. This agent receives the original query, execution trace summary, and all tool-generated data to produce the final human-readable response. The entire execution process is streamed via Server-Sent Events (SSE) technology, providing users with real-time progress feedback and significantly optimizing human-machine interaction experience.

\subsection{Interactive Module}
As the Unified User Interaction Exit, the Interactive Summary module evaluates the alignment between the final results and the user's original query while supporting compensatory retry mechanisms. In output generation, the module supports diverse output formats and styles (e.g., markdown tables, JSON summaries, narrative reports), enhancing the system's intelligent attributes.

\section{Experiment}
To rigorously validate the effectiveness of the proposed Z-Space: A Multi-Agent Collaborative Tool Orchestration Framework for Enterprise-Grade LLM Automation, this study conducts comprehensive comparative experiments across multiple metrics:

1. Accuracy: Measures the alignment between system outputs and ground-truth requirements.

2. Token Consumption: Evaluates computational efficiency through LLM inference token costs.

3. Planning Success Rate: Quantifies the robustness of task execution workflows.

Additionally, the performance of the FSWW algorithm is visualized through semantic vector field diagrams and attention heatmaps. The following sections detail the experimental design, implementation methodology, and quantitative analysis.

\subsection{Experiment Design}
\subsubsection{Dataset Construction}
This study builds a dataset based on enterprise-grade tool invocation scenarios, comprising 541 KBT tools and 280 user instructions:

Tool Dataset. The tool specification model integrates manually annotated fields (name, description, execution environment) and AI-generated fields (summary, entity tags, capability tags), covering multi-dimensional semantic information (as shown in figure \ref{Tool Model Design}). To support the proposed Z-Space framework for data generation-oriented multi-agent collaborative tool invocation and enable comprehensive evaluation, an enterprise-scale tool dataset was constructed. In enterprise data generation scenarios, the MCP concept is largely equivalent to the tool concept, resulting in deployments containing dozens to hundreds of tools within a single MCP server.

\begin{table}[H]
 \caption{Tool Model Design}
  \centering
  \begin{tabular}{ccc}
    \toprule
    Key     & Description     & Generation Method \\
    \midrule
    name & name of tool  & manual     \\
    description     & functions and parameters description of tool & manual     \\
    environment     & execution environment       & manual  \\
    summary & summary of tool-related information & AI \\
    entityTags & entities associated with tool execution  & AI \\
    capabilityTags  & capability tags of the tool &  AI \\
    \bottomrule
  \end{tabular}
  \label{Tool Model Design}
\end{table}

User Instruction Set. Instructions are categorized by execution steps: single-step (40\%), dual-step (30\%), and multi-step (30\%), covering business scenarios such as data generation and querying.

\begin{table}[h]
 \caption{User Instruction Set}
  \centering
  \begin{tabular}{ccccccc}
    \toprule
    Number of Execution Steps & 1 & 2 & 3 & 4 & 5 & 6 \\
    \midrule
    Proportion(\%) & 40  & 30 & 15 & 10 & 3 & 2     \\
    \bottomrule
  \end{tabular}
  \label{tab:table1}
\end{table}

This study compares Z-Space with two mainstream frameworks through systematic comparative experiments. All primary agents adopt Qwen-plus and Qwen-max LLMs to eliminate performance variance from different LLMs.

\paragraph{LLM Baseline.} Architecture: Identical Z-Space multi-agent structure with intent parsing, reasoning execution, and summary interaction agents. Workflow: The parsed intent and all tool information are fed as context into Qwen-plus for direct intent parsing and tool selection.

\paragraph{LLM+RAG Baseline.}: Architecture: Z-Space multi-agent framework augmented with standard RAG pipeline.Workflow: Intent Parsing: User query undergoes intent recognition.
RAG Tool Retrieval: Standard RAG process filters Top5 candidate tools.
Execution: Selected tools are executed by the reasoning execution agent.
Output Generation: Final response is generated by the summary interaction agent.

\subsection{Experiment Analysis}
\subsubsection{Accuracy Comparison Experiment}
This study conducts comparative experiments on the self-constructed dataset and instruction set, benchmarking Z-Space against baseline models. The results demonstrate that Z-Space achieves significant improvements in both accuracy and token efficiency. Through systematic parameter control experiments, this study determined the optimal hyperparameter configuration as follows: $\alpha$=0.5, $\beta$=0.1, $\gamma$=0.6, $\lambda$=0.6, $\epsilon$=0.001, with a correlation coefficient threshold $\Delta$ set at 0.9. All subsequent experiments adopted this standardized parameter configuration.

\begin{table}[H]
 \caption{Accuracy Comparison Experiment}
  \centering
  \begin{tabular}{ccc}
    \toprule
    Method     & Accuracy     & Token Consumption \\
    \midrule
    LLM &  27.32 & 6962.44     \\
    LLM + RAG     & 82.95 & 242.54     \\
    Z-Space     & 92.00       & 260.37  \\
    \bottomrule
  \end{tabular}
  \label{tab:table2}
\end{table}

Z-Space achieves 92\% tool execution accuracy, surpassing the traditional single-LLM framework by 64.68\% and outperforming the LLM+RAG architecture by 9.05\%. This validates the effectiveness of Z-Space's multi-agent architecture (intent parsing + reasoning execution) and the FSWW algorithm in enhancing semantic information.

Additionally, Z-Space reduces token consumption by 96.26\% compared to the LLM baseline. This efficiency gain stems from the RAG-based filtering mechanism, which selectively identifies tools semantically relevant to user intent from the repository. By avoiding full-context injection, the framework prevents token inflation and attention dilution in LLMs, thereby improving large model utilization efficiency.

\begin{figure}[H]
    \centering
    \includegraphics[width=0.75\linewidth]{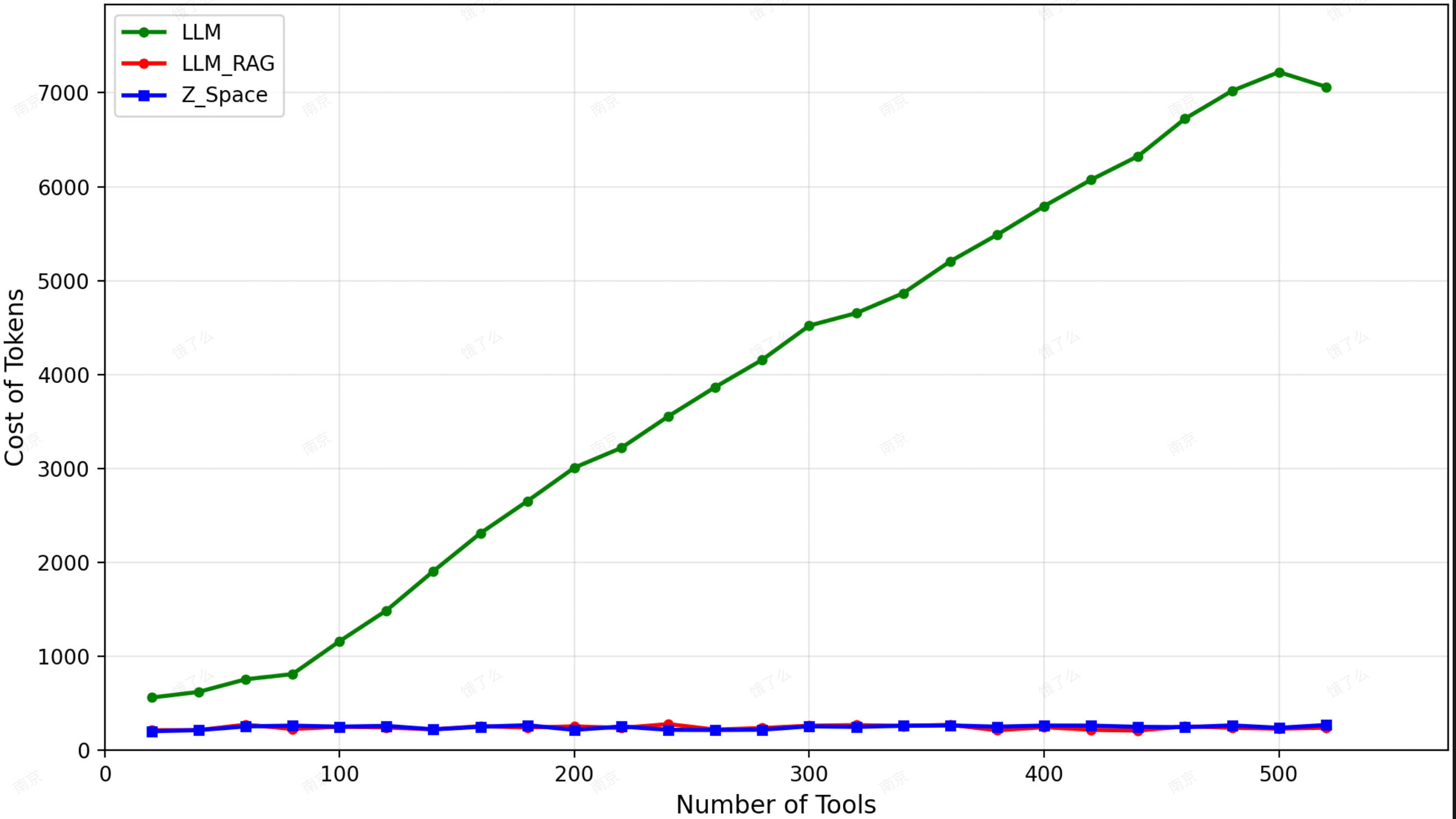}
    \caption{Tokens Cost with Number of Tools}
    \label{kzxsy}
\end{figure}

\paragraph{Scalability Comparison Experiment with Increasing Tool Counts}
To evaluate Z-Space's scalability, we gradually increased the tool count from 20 to 520 in the test environment and measured token consumption during tool retrieval. The experimental results (shown in figure \ref{kzxsy}) demonstrate that the LLM baseline exhibits linear growth in token consumption as the tool repository size increases. In contrast, LLM-RAG and Z-Space maintain stable token consumption during reasoning execution because they retrieve only 4–6 tools per query, showcasing superior scalability and resource stability. Furthermore, although the new FSWW semantic enhancement algorithm was introduced, it did not significantly increase token consumption or impose additional model costs.

\paragraph{Multi-Step Instruction Task Experiment}
This study also conducted experiments on instructions with varying execution steps. Multi-step instructions require sequential invocation of multiple tools to complete, a common scenario in test data generation. For example: "Create a test product, place an order using account 123456, and advance the order status to pending payment"—this instruction sequentially requires four tools: test product creation, user information query, order creation, and order status advancement.

\begin{figure}[H]
    \centering
    \includegraphics[width=0.75\linewidth]{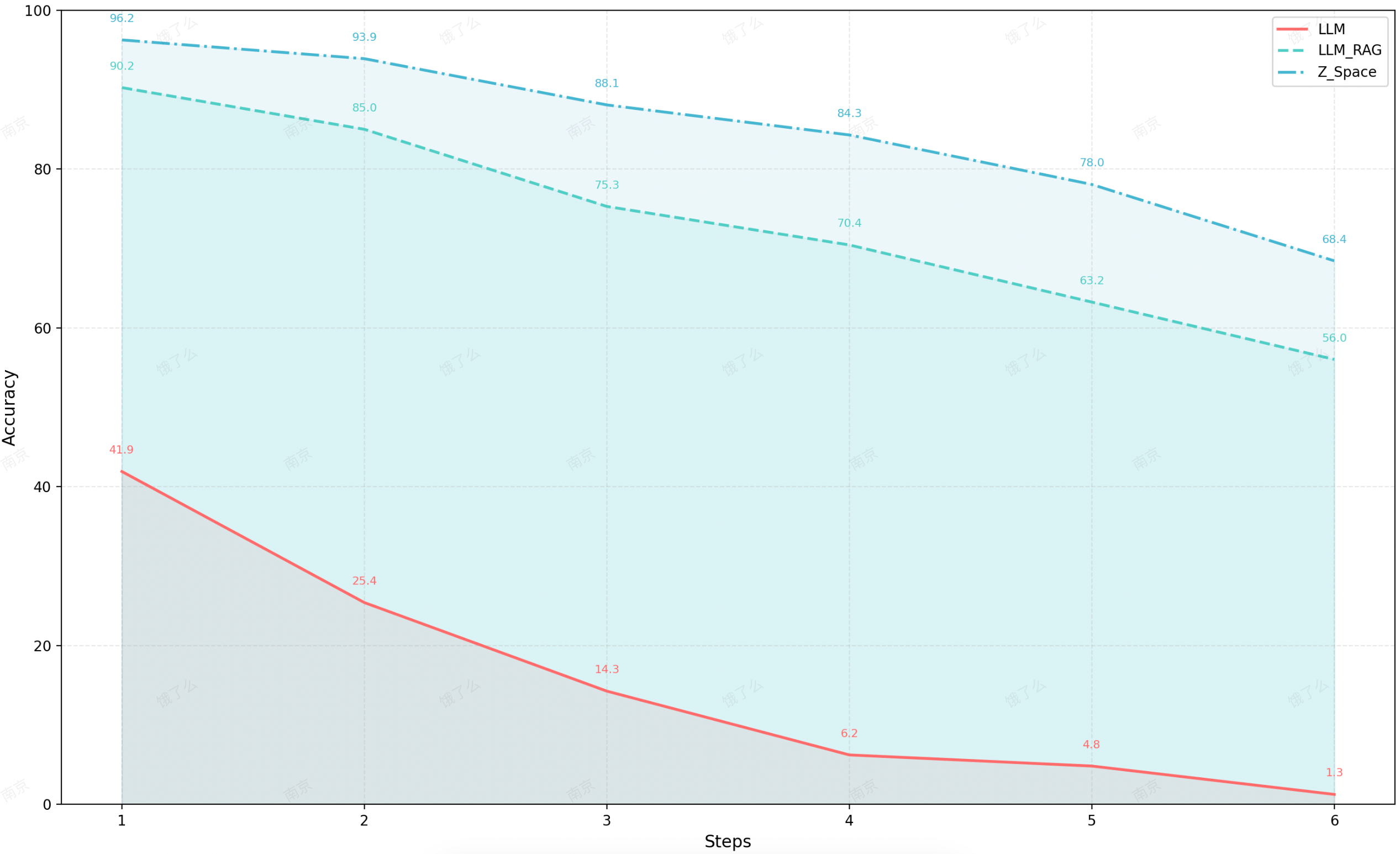}
    \caption{Model Performance Decay Across Steps}
    \label{dbzsy}
\end{figure}

Experiments were systematically conducted on instructions with different step counts. The results (shown in figure \ref{dbzsy}) reveal a declining trend in tool execution accuracy as the number of required steps increases. The LLM framework exhibits severe performance degradation in multi-step tasks: it achieves 41.9\% single-tool accuracy for single-step instructions, but accuracy plummets to 1.3\% at Step=6 (a 40.6\% decline). This highlights that relying solely on multi-agent architecture leads to critical failures in state tracking and error accumulation during complex tool orchestration, undermining logical consistency in long-chain tasks.

While LLM-RAG improves initial task performance (90.2\% accuracy at Step=1) through RAG mechanisms, its accuracy still shows progressive decay with increasing step counts, dropping to 56.0\% at Step=6. This indicates that traditional RAG, though effective in improving single-step tool accuracy, has limited capability in capturing semantic dependencies across multi-step user intents and tool interactions.

In contrast, Z-Space demonstrates superior performance stability. It achieves the highest initial accuracy (96.2\% at Step=1) and the slowest degradation rate, maintaining 68.4\% accuracy at Step=6 (only a 27.8 percentage-point decline). This proves that Z-Space's integration of RAG mechanisms and FSWW algorithms effectively strengthens core semantic information in user intents and tool descriptions, significantly enhancing robustness in long-chain reasoning tasks.

\subsection{Visual Demonstration of FSWW Algorithm Outputs}
This experiment used the UMAP dimensionality reduction algorithm to visualize 100 tool executions in a 3D semantic space, comparing the impact of the conventional RAG framework and the FSWW-enhanced framework on the semantic embedding distributions of execution plans (Execution Plan) and tools (Tool). 

\begin{figure}[ht]
    \centering
    \begin{minipage}[b]{0.45\textwidth}
        \centering
        \includegraphics[width=\textwidth]{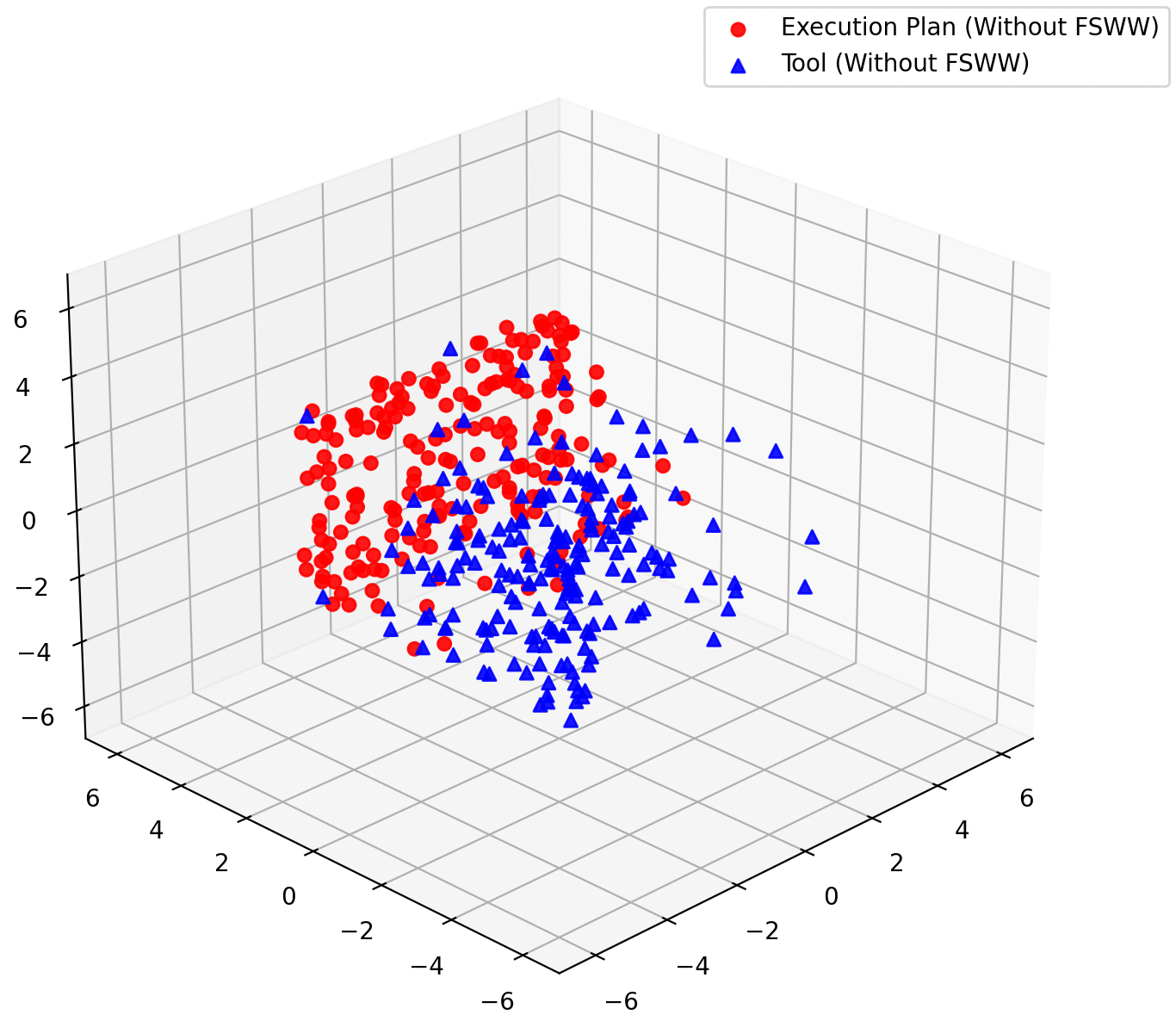} 
        \caption{Result without FSWW}
        \label{left}
    \end{minipage}
    \hfill
    \begin{minipage}[b]{0.45\textwidth}
        \centering
        \includegraphics[width=\textwidth]{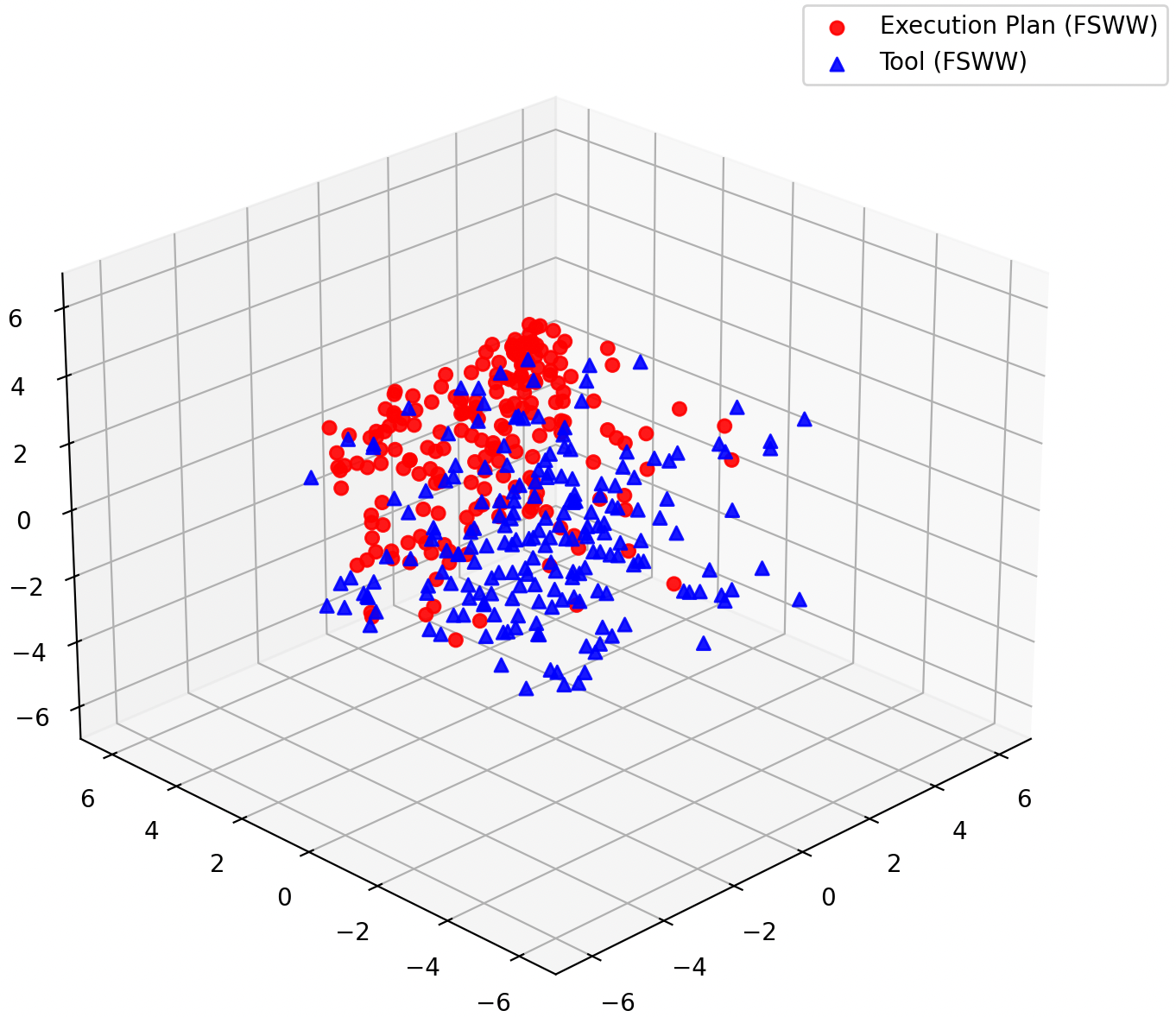} 
        \caption{Result with FSWW}
        \label{right}
    \end{minipage}
\end{figure}

The figure \ref{left} shows the semantic space of the conventional RAG framework without FSWW integration: red circles (Execution Plan) and blue triangles (Tool) exhibit a highly dispersed distribution pattern. These two data clusters show significant overlap in the 3D space, with outliers widely scattered in extreme regions of the coordinate axes. This indicates that the conventional RAG framework suffers from poor semantic separability between execution plans and tools, where large embedding distances between intents and tools constrain matching accuracy in multi-step tasks and introduce semantic ambiguities.

The figure \ref{right} demonstrates the optimization effects of FSWW technology: through weighted subspace projection, tool vectors are dynamically pulled toward semantic directions dominated by keywords, causing the point cloud distributions of Execution Plan and Tool to converge significantly. The improvements include enhanced clustering compactness (both data clusters form high-density regions in the central area with reduced internal dispersion), boundary clarification (overlapping regions decrease by ~40\% with clearer semantic boundaries between execution plans and tools), and a distance shortening effect (the average embedding distance between intents and tools drops from 2.82 to 1.09 in 3D space). This validates that FSWW effectively narrows the semantic gap between task intents and tool functionalities through its alignment mechanism. The comparison directly proves that FSWW's feature reweighting and subspace projection significantly improve the geometric structure of the semantic space, providing a more robust foundation for multi-step tool invocation frameworks.

\section{Conclusion}
This study addresses the challenges of intent understanding and semantic matching in enterprise-grade data generation scenarios with large-scale tool repositories by proposing the Z-Space multi-agent collaborative tool invocation framework. The core contribution lies in constructing a closed-loop decision-making system from semantic understanding to physical execution. It first employs intent recognition agents to parse unstructured natural language instructions into structured semantic graphs, capturing deep user intents. Then, the innovative Fused Subspace Weighting Algorithm (FSWW) plays a critical role in tool filtering by simulating attention and residual mechanisms from deep learning. This algorithm enhances fine-grained semantic alignment between intents and tool descriptions without model fine-tuning, effectively resolving matching biases in complex semantic scenarios like test data generation where traditional RAG methods struggle. Finally, reasoning execution agents orchestrate asynchronous tasks based on intent trees, ensuring fault tolerance and high concurrency efficiency for multi-step operations. Experiments demonstrate that Z-Space achieves dual breakthroughs in efficiency and reliability for enterprise-grade LLM automation: it improves tool invocation accuracy to 92\% while reducing reasoning costs by 96.26\% in complex data generation scenarios like Eleme.

\bibliographystyle{unsrt}  


\end{document}